\documentstyle[epsfig]{mn}
\newcommand{\samethanks}
        {{\Huge $^\star$}}
\newcommand{\etal}
	{et al.}
\newcommand{\eg}
	{e.g.}
\newcommand{\cf}
	{cf.}
\newcommand{\ie}
	{i.e.}
\newcommand{\bj}
        {b_{\rm J}}

\begin{document}

\title[Quasars in the 2dFGRS]
{Spectroscopic detection of quasars in the 2dF Galaxy Redshift Survey}

\author[Darren S.\ Madgwick et al.\ ]
       {Darren S.\ Madgwick$^{1}$\thanks{
                E-mail: dsm@ast.cam.ac.uk (DSM);
                phewett@ast.cam.ac.uk (PCH);
                mortlock@ast.cam.ac.uk (DJM);
                lahav@ast.cam.ac.uk (OL)},
        Paul C.\ Hewett$^{1}$\samethanks ,
        Daniel J.\ Mortlock$^{1,2}$\samethanks
\newauthor
        and
        Ofer Lahav$^{1}$\samethanks \\
        $^1$Institute of Astronomy, Madingley Road, Cambridge
        CB3 0HA, U.K. \\
        $^2$Astrophysics Group, Cavendish Laboratory, Madingley Road,
        Cambridge CB3 0HE, U.K.}

\date{
Accepted 2002 March 19. 
Received 2001 November 23; 
in original form 2001 October 26.}

\pagerange{\pageref{firstpage}--\pageref{lastpage}}
\pubyear{2001}

\label{firstpage}

\maketitle

\begin{abstract}

The 100,000 spectra from the 2 degree Field Galaxy Redshift Survey
(2dFGRS) in the 100k Public Data Release represent the largest single
compilation of galaxy spectra available. By virtue of its sheer size
and the properties of the photometric catalogue that defines the
sample, the 2dFGRS is expected to contain a number of potentially
interesting objects other than galaxies. A search of the spectra in the 100k
Data Release results in a census of 55 candidate high-redshift ($z\ge
0.3$) quasars.  
One additional 2dFGRS spectrum of a low-redshift
galaxy shows an apparent anomalous broad emission feature perhaps
indicating the presence of a gravitationally lensed quasar.  
These objects have been identified primarily using two
automated routines that we have developed specifically for this task,
one of which uses a matched filter and the other a wavelet transform.
A number of
the quasar images possess complicated morphologies, suggesting the
presence of either diffuse foreground objects along the line--of--sight
or very nearby point sources. The quasar catalogue will form a target list
for future absorption and lensing studies, as well as providing an
assessment of the loss of quasars with non-stellar images from the
companion 2dF QSO Redshift Survey.

\end{abstract}


\section{Introduction}
\label{section:intro}

Very large surveys are playing an increasingly important role in the
progress of observational astronomy, with all-sky photometric and
spectroscopic data--sets particularly prominent. The most advanced surveys
in the latter category are the 2dF Galaxy Redshift Survey
(2dFGRS; Colless \etal\ 2001) and the Sloan Digital Sky Survey (SDSS;
York \etal\ 2000), both of which already include well over $10^5$
spectra. A feature of many of the most ambitious surveys is the
incorporation of the rapid release of survey data to the astronomical
community, together with user-friendly tools and interfaces that allow
exploitation of this resource. Indeed the 2dFGRS's release of $10^5$
spectra\footnote{{\tt http://www.mso.anu.edu.au/2dFGRS/}} is the largest
catalogue of galaxy spectra available.

The 2dFGRS 100k Data Release has already formed the basis for
investigations key to the primary goals of the survey, including
constraints on the luminosity function of local galaxies (Sadler
\etal\ 2001; Magliocchetti \etal\ 2002; Madgwick \etal\ 2002) and their
correlation properties (Peacock \etal\ 2001; Percival \etal\ 2002;
Norberg \etal\ 2001), as well as studies of the local cluster population (de
Propris \etal\ 2001).  Comparable results have also been obtained from
the SDSS (e.g. Blanton \etal\ 2001; Connolly \etal\ 2002).

An additional benefit of these large databases is the presence of new
and unusual objects, albeit in small numbers.  The SDSS has already
discovered several unusual dwarf stars (Strauss \etal\ 1999; Harris
\etal\ 2001) and a large proportion of the highest-redshift quasars
now known (Fan \etal\ 2001).  The morphological selection involved
in the definition of the 2dFGRS targets reduces the heterogeneity of
the constituent object populations compared to the SDSS but the 100k
Data Release should still include a variety of interesting objects.

In this paper we focus on the population of broad line quasars and
active galactic nuclei (AGN) with redshifts essentially beyond the
limit of the number--redshift selection function for the
normal galaxies that dominate the sample. Such a census of quasars and AGN
with redshifts $z \ga 0.3$ has a number of applications. 
Firstly, many of the quasars with
redshifts $z \ga 0.5$ in the 2dFGRS will lie close in projection to lower
redshift galaxies -- the `composite' quasar plus galaxy image
having been classified as non-stellar. These quasar--galaxy pairs
provide the opportunity to probe the physical conditions in the
interstellar medium of the galaxies via absorption line studies of the
background quasars.  A smaller fraction of quasars, concentrated
particularly at lower redshifts, $z \la 0.5$, will show evidence for
the presence of host galaxies. The sub-sample of such objects will
include quasars with some of the most luminous hosts known. 
Secondly, a
few objects may represent new examples of strong gravitational lensing,
adding to the still very small number of multiply imaged quasars
known. 
Finally, as
the companion 2dF QSO Redshift Survey (2QZ; \eg\ Croom \etal\ 2001)
photometric catalogue deliberately includes only unresolved sources,
the frequency and properties of quasars present in the 2dFGRS provides
information relating to the completeness of the 2QZ as a function of
image morphology and redshift.  

Following a brief description of the 2dFGRS data
(Section~\ref{section:2dfgrs}) the search methods employed are detailed
in Section~\ref{section:search}. Issues related to candidate selection
are discussed in Section~\ref{section:identification} and the resultant
object catalogue is presented in Section~\ref{section:catalogue},
together with comments on individual objects. The paper concludes with
a brief discussion of the potential uses of the quasar sample in
Sections~\ref{section:discussion} and \ref{section:conclusion}.

\section{The 2\lowercase{d}FGRS}
\label{section:2dfgrs}

The 2dFGRS is, formally, a spectroscopic survey of $\sim 2.5 \times
10^5$ apparently non-stellar objects to an isophotal magnitude 
limit of $\bj = 19.45$.
Our search for quasars and broad line AGN is confined to the first
subset of $10^5$ spectra made publicly available as the 100k Data
Release (Colless \etal\ 2001).

The 2dFGRS spectra are obtained using the 2dF instrument on the
Anglo-Australian Telescope (AAT) and represent $\sim 45$-minute
integrations through fibres with an angular diameter on the sky of
$\sim 2.1\,$arcsec (see Lewis \etal, 2002). The wavelength 
coverage extends from approximately
$3700$\AA\ to $8100$\AA. Strong night-sky emission lines, notably
\hbox{O\,{\sc i}} 5577\AA, \hbox{Na\,{\sc i}} 5890\AA\ and \hbox{O\,{\sc i}}
6300\AA, effectively
produce $\sim 50$\AA-wide gaps in the spectra of all except the
brightest objects.  The nominal signal--to--noise (S/N) ratio for a
continuum-dominated object at the survey limit is $\sim 10$ per
10\AA\ bin, although many of the spectra are of significantly
lower quality.

The majority of the targets are local galaxies, with an average
redshift of $\langle z \rangle =
0.1$, although there is a $\sim 5$ per cent `contamination' by
Galactic stars. Morphological classification of the photometric input
catalogue was taken from the Automatic Plate Measuring (APM) survey
(Maddox \etal\ 1990) of United Kingdom Schmidt Telescope (UKST)
photographic plates.  The star--galaxy separation algorithm, described
in detail by Maddox \etal\ (1990), is highly effective when
applied to isolated images, and able to select galaxies with an
efficiency (\ie\ the fraction of selected objects which are
galaxies) of $\sim 97$ per cent.  Classification of `composite' images,
where the isophotal boundaries of two or more images overlap and the
APM parameterises the resulting composite image, present a more
difficult problem given the limited number of moment-based parameters
provided by the APM to describe each image. 
For objects with magnitudes within the range included in the 2dFGRS
essentially all close pairs of objects with angular separations $\la
8\,$arcsec (depending on the quality of the plate material) are
detected as composite images. The majority of such images have been
eliminated from the 2dFGRS spectroscopic target catalogue through a
selection according to image classification parameter and direct visual
inspection (Colless \etal\ 2001; Section 5.4). However, a small fraction
of star--galaxy and even star--star pairs are included in the
2dFGRS target list and such objects are responsible for the bulk of the
contamination by Galactic stars.

The population of unresolved quasars and AGN is subject to exactly the
same image blending that can occur with Galactic stars, but the surface
density on the sky is much smaller and hence the predicted frequency in
the 2dFGRS survey correspondingly lower. Nonetheless, adopting a surface
density of $10\,$deg$^{-2}$ for quasars with magnitudes $b_{\rm J} \la 19.5$,
simple geometric arguments predict that the 2dFGRS 100k Data Release should
contain several tens of quasars.

\section{Search Methodologies}
\label{section:search}

Our search strategy involved a combination of visual inspection
(Section~3.1), confined to the most extreme objects in the 100k Data
Release, and the application of two automated routines (Sections~3.2
and 3.3).  The combined results are presented in
Section~\ref{section:identification}.

\subsection{Visual inspection}

Our starting point involved a simple query of the Data Release,
selecting all objects with heliocentric redshifts $z_{\rm helio} \ge
0.3$.  All 422 spectra satisfying the $z_{\rm helio} \ge 0.3$
constraint were inspected visually to identify objects with evidence of
broad emission lines.

\subsection{Matched filter-based search}

The first of our automated routines adopts a `conventional' approach
to filtering the spectra before applying a matched filter detection
procedure.  A mask was defined to exclude pixels with wavelengths
within the intervals $5557$--$5597$\AA, $5870$--$5910$\AA,
$6280$--$6320$\AA, $6340$--$6380$\AA, $6850$--$6950$\AA,
$7150$--$7350$\AA, $7580$--$7700$\AA\ and $7880$--$8500$\AA. These
regions are dominated by either residual
sky-subtraction errors in the vicinity of strong night-sky emission
lines, \eg\ \hbox{O\,{\sc i}} $\lambda$5577, NaD $\lambda$5890, or
suffer from strong absorption within the atmosphere or the 2dF fibres
themselves. Pixels within $40$\AA \ of the nominal start and end
wavelengths of the spectra were also excluded to avoid a number of
artifacts arising from the data reduction or the instrument.

A median filter with a box-length of 201 pixels ($\simeq 840$\AA) was
then applied to the spectra. The resulting `continuum' spectra retain
any discontinuities present in the original spectra, \eg\ the
$4000$\AA-break in many of the galaxies, but the presence of
absorption and emission features with widths $\la 400$\AA \ is greatly
reduced. The continuum spectra were then subtracted from the original
spectra to produce `difference' spectra. Further discussion and
examples of the effect of median filters applied to low-resolution
astronomical spectra can be found in Hewett \etal\  (1985).

The `matched filter' was derived from the Francis \etal\ (1991)
composite quasar spectrum, extended using additional data to
$7500$\AA.  A low-order spline fit to the quasar continuum was
subtracted from the composite spectrum to leave a residual spectrum
dominated by prominent broad emission lines.

For each difference spectrum the template was redshifted and
interpolated onto the same wavelength array. The template was then
multiplied by an approximation to the wavelength-dependent sensitivity
of the 2dFGRS observations. A variance array is provided for each
2dFGRS spectrum in the 100k Data Release and a straightforward
cross--correlation between the difference spectrum and template
spectrum was then performed to give a signal--to--noise ratio (Hewett
\etal\ 1985; Equation 3). The template was incremented by $\delta z =
0.01$ and the search performed over the redshift range $0 \le z \le
3.5$. The associated scale factor for the template spectrum is readily
calculated to give the amplitude of the signal (Equation
2 of Hewett \etal\ 1985). 
A goodness--of--fit estimate was calculated from the fractional
reduction in $\chi^2$ between the difference spectrum alone and the
difference spectrum after the subtraction of the scaled template.
Thus, for each spectrum, `detections', above some specified
S/N-threshold, together with associated goodness--of--fit measures and
redshifts are generated.

\subsection{Wavelet search}

In addition to the matched filter search we have also conducted an
automated search of the spectra using a wavelet transform-based
routine. 

The distinguishing feature of quasars in spectra of low
S/N ratio is the presence of broad emission lines.
Wavelet transforms allow the separation of a spectrum into different
(Fourier) frequency domains whilst preserving the spatial information
(\ie\ the emission line positions).  This allows us to easily separate
out broad from narrow features in each spectrum and hence make the
identification of quasar emission signals more straight-forward. The
key is the use of a localised 
filter kernel, the width of which varies according to the resolution
being sampled. This contrasts with a conventional Fourier transform
where the filter kernel is not localised and the spatial information is
obscured during the transformation.

There are a variety of different implementations of the wavelet
transform (\eg\ Mallat 1998), each of which has its own particular
benefits. For the analysis of the 2dFGRS the {\em $\acute{a}$ trous} 
(`with holes')
algorithm (Starck, Siebenmorgen \& Gredel 1997 and references
therein) was used  
because it has a particularly simple inverse, making the transformed
spectra easy to interpret.

\begin{figure}
 \psfig{figure=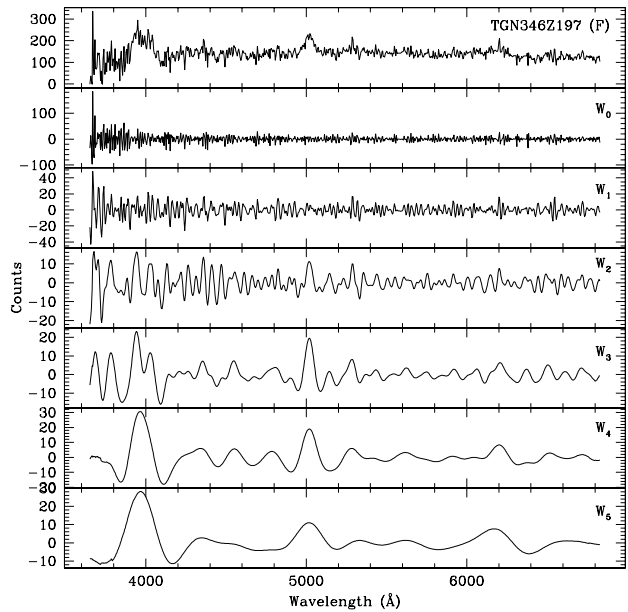,width=3.5in}
 \caption{The original spectrum of TGN346Z197 is shown in the top
 panel.  The remaining panels show each successive resolution domain
 of the wavelet transform $W_k$ of this spectrum.  It can be seen that
 the broad quasar emission features are well isolated in the last
 two domains and that most of the noise is restricted to the first few.}
 \label{exfig}
\end{figure}

Using this transformation yields a set of different resolution
domains $W_k$ for any given spectrum.
An example of this is shown for a 2dF spectrum
in Fig.~\ref{exfig}.
The kernel we have employed is a normalised Gaussian sampled at five
equally spaced points. 
This process can be carried on for any number of steps $k$.  In the
case of the 2dFGRS galaxy spectra we find that there is no additional 
emission information left after about 7 or 8 steps.

\subsubsection{Detection}

As illustrated in Fig.~\ref{exfig}, the effect of the wavelet
transform is to localise the broad emission 
line signature of any quasar in the low-resolution domains, which are
free from most of the pixel--to--pixel noise and any spurious
high frequency features that may be present.

Quasars can now be identified by cross-correlating these low-frequency
resolution domains $W_k$ with the corresponding domains $W'_k$ of the 
transformed quasar template.  This process is carried out in a similar
fashion to the matched filter method, but with a few minor modifications. 

Empirically, the most effective resolution domains to sample for quasar
lines are $W_3$, $W_4$ and $W_5$. Because the wavelet transform of the
quasar template is performed before redshifting, the width of the
quasar features in a given resolution domain will increase with
increasing redshift, introducing a bias against detecting the more
distant quasars. For this reason two sets of cross-correlations
were
performed. In the first, domains $W_3$ and $W_4$  of the 2dF
spectrum were
cross-correlated with $W_3$ and $W_4$ of the quasar template (to detect
low-$z$ objects). In the second, domains $W_4$ and $W_5$ of the 2dF spectrum
were cross-correlated with $W_3$ and $W_4$ of the quasar template (to detect
high-$z$ objects). 
The combined cross-correlation is then the sum of the four individual
cross-correlations. 

The goodness--of--fit for any quasar identifications is then
determined as described for the matched--filter search.

\section{Identification of high-redshift objects}
\label{section:identification}

\subsection{Visual inspection}

The 2dFGRS redshift-assignment for spectra in the survey has been
undertaken using techniques that are highly effective for galaxies with
absorption line or emission line spectra produced by stellar
photospheres and \hbox{H\,{\sc ii}} regions.  
Unsurprisingly, however, objects whose
spectra are dominated by broad emission lines and/or possess redshifts
$z \ga 0.4$ are not well suited to redshift determinations using
template spectra of normal galaxies.  Only 150 of the 422 spectra with
$z_{\rm helio} \ge 0.3$ have redshift assignments of high confidence
(quality = 4 or 5) and it is apparent that a number of the spectra have
incorrect redshift assignments.  Objects with erroneous redshifts
include some spectra of high S/N ratio and some objects
with high confidence redshift assignments. The limitations of the
2dFGRS redshift determination procedure are essentially irrelevant to
the primary goals of the survey but care should be exercised
when utilising this very small number of extreme objects
from the 100k Data Release. The sample identified from the visual
inspection consisted of 37 quasars and AGN with redshifts $z_{\rm
helio} \ge 0.3$.

\subsection{Automated searches}

Both the automated routines described in the previous section were
highly effective in identifying spectra that 
show broad emission lines. 
However, the limiting factor for the automated routines is the large
number of false-positive detections that arise due to a wide variety
of features and artifacts present in the 2dFGRS spectra.

A full description of the properties of the 2dFGRS spectra is beyond
the scope of this paper, and the majority of the phenomena that affect a
search for rare objects are described in the 2dF instrument on-line
documentation\footnote{\tt http://www.aao.gov.au/2df/}.  However,
mention of some of the features in the context of searches for quasars
is appropriate.

\begin{figure}
\psfig{figure=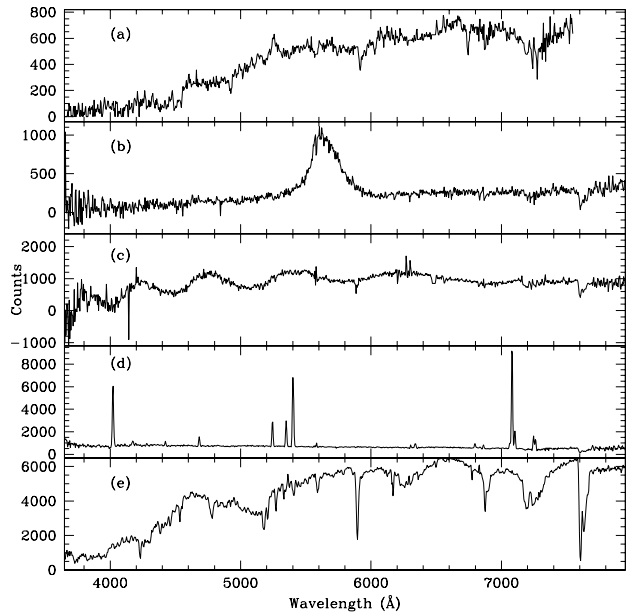,width=3.5in}
\caption{Examples of spurious detections from the automated routines.
From top to bottom: (a) unknown artifact at $5200$\AA, 
(b) LED contamination, (c) fibre ringing, (d) very strong narrow line 
emission and (e) high S/N spectrum of a Galactic star showing
strong absorption features.}
 \label{badfig}
\end{figure}

A variety of instrument-related phenomena can result in spurious
detections.  Prior to August 1999 the second 2dF spectrograph fed an
engineering-grade CCD that possessed a significant number of cosmetic
defects. The majority of the artifacts not removed during the
data reduction procedure involve relatively narrow features and/or
sudden discontinuities. The features themselves do not in
general provide good mimics for broad emission lines but their presence
can on occasion cause the filtering routines to behave unpredictably,
resulting in numerous spurious detections. A number of spectra
obtained through fibres 39--50 possess an apparent broad emission
feature at a wavelength of $\sim 5200$\AA \
(Fig.~\ref{badfig}a). Examples include 
objects TGN403Z028, TGN420Z090, TGS181Z082, TGS181Z095 and TGS202Z267.

Contamination from light emitting diodes (LEDs), particularly broad
features at $\sim 6600$\AA \ and $\sim 7100$\AA, creates a number of
`emission' features in some spectra acquired before about mid-1999.
These features are often, quite correctly, identified as candidate
lines by the automated routines. In addition a number of fibres are 
particularly
prone to optical ringing which results in a strong modulation of the
object spectrum that frequently results in spurious detections.  
It is believed that the ringing is due to an interference fringing
effect resulting from small cracks at the input ends of the optical
fibres (Lewis \etal 2002).  Examples of both LED contamination and
optical ringing are shown in Fig.~\ref{badfig}(b) and (c).

At low redshifts the very large
number of spectra with narrow emission lines, the majority of which
result from processes associated with star-formation, produce a
large number of false detections (Fig.~\ref{badfig}d). The low
S/N of the majority of the spectra means that establishing the presence
of broad wings to emission lines is not possible. At higher redshifts
the difficulty is reduced dramatically as only a small number of
star-forming galaxies are bright enough to be included within the
2dFGRS flux-limited sample. Indeed, we identified only 3 objects with
narrow emission line spectra and redshifts $z \ge 0.4$. 

Bright stars can result in very high S/N spectra
(Fig.~\ref{badfig}e) and filtering techniques that do not incorporate
knowledge of stellar spectral energy distributions almost invariably
leave apparent broad emission and absorption signatures in the filtered
spectra. 

The large number of false-positive detections resulting from the range
of phenomena described above mean that the automated techniques are
limited more by the swamping of the small number of real quasars than
by the question of whether a detection at a given significance is real.
Following some experimentation, candidates from the wavelet-based
search were selected according to a fractional reduction in
the goodness--of--fit of 
$\delta \chi^2 \ge 0.25$, producing 711 objects. The cut-off values
adopted for the matched filter-based search were $S/N\ge2.5$ and
$\delta \chi^2 \ge 0.2$ producing 1277 candidates. 
 
The candidate spectra were then inspected visually to eliminate the
spurious detections that constitute the bulk of the candidates.
Encouragingly, the automated techniques independently recover the
majority of objects identified from the visual inspection of the
$z_{\rm helio} \ge 0.3$ sample and the majority of candidate quasars are
common to all three lists. The combined sample of candidate quasars
consists of 55 objects, of which 37, 49 and 47 were identified in the
visual, matched filter and wavelet searches respectively. 

\subsection{Observer Comments}

To verify that the combination of visual inspection and the automated
search was effective in identifying candidate quasars the `observer
comments' ({\tt{Z\_COMM}}, Colless \etal\ 2001) were searched for
references to possible quasar lines or otherwise unusual spectra. Of
124 such references, 29 were included in our candidate list and visual
inspection of the remainder revealed no additional candidates.

\section{High-redshift objects in the 100\lowercase{k} Data Release}
\label{section:catalogue}

\subsection{The Object catalogue}
A total of 55 candidate quasars with redshifts $z \ge 0.3$ were
identified using one or more of the three methods described in 
Section~\ref{section:identification}. Table 1 provides summary
information for these objects. The 2dFGRS
name, celestial coordinates and $b_{\rm J(2dF)}$ magnitudes are all taken
directly from the 100k Data Release. The quasar redshifts, $z_{\rm qso}$,
are derived via cross-correlation of the quasar template (Section 3)
with the individual spectra. The probability a candidate represents a
quasar is quantified via a simple integer quality flag, $ID$:  
\begin{enumerate}
\item $ID=1$, {\em Possible
quasar} -- spectrum contains only one broad emission feature
detected at low S/N (5 objects);
\item $ID=2$, {\em
Probable quasar} -- spectrum contains only one broad emission
feature detected at moderate S/N, or spectrum contains two or more
broad emission features detected at low S/N (9 objects);
\item $ID=3$,
{\em Definite quasar} -- spectrum contains two or more broad
emission features detected at moderate or high S/N (41 objects). 
\end{enumerate}
A high degree of confidence can be placed in the
candidates with $ID=3$, while candidates with $ID=1$ certainly require
further spectroscopic observations to clarify the reality of the
putatitive broad emission features.

Figure~\ref{sample} shows the 2dFGRS spectra together
with the redshifted quasar template used to make the identification.
Interpolation has been performed over strong sky emission lines and a
few spurious features masked out.  Also shown are $30 \times
30$ arcsec postage stamp images, in conventional astronomical orientation,
taken from the SuperCOSMOS Sky Survey (SSS; Hambly et al. 2001) scans of the
UKST $\bj$ survey plates.  The scale of these postage stamp
images is $0.7''$ per pixel.

\subsection{Corrected magnitudes}

The 2dFGRS $\bj$ magnitudes are calculated assuming that
the images do not contain unresolved point sources. However, the
catalogue of candidate quasars includes objects that are only
marginally resolved as well as objects that consist of two or more
distinct components, at least one of which is unresolved. An automated
scan of a UKST photographic plate shows a
tight relation between the integrated brightness and the peak
brightness of unresolved images. The relation arises because of the
very similar profiles of the unresolved sources and the peak brightness
of unresolved images with $\bj \la 20.5$ exceeds that of the majority
of galaxies in the 2dFGRS. As a consequence, the peak brightness of
marginally resolved or composite images, such as those of the candidate
quasars, can be used to estimate the brightness of any unresolved
component. The SSS has been used to
provide magnitude estimates for the unresolved components present in
each of the candidate quasar images. The SSS catalogues were
integrated to provide the SuperCOSMOS $\bj$ magnitudes and peak
intensities ($I_{\rm PEAK}$) for all images classified as `stellar'
within a $15\,$arcmin radius of each candidate quasar. A low-order
fit to the well-defined $\bj$ vs $I_{\rm PEAK}$ relation was performed for
objects surrounding each candidate and $\bj$ for the candidate quasar,
based on the value of $I_{\rm PEAK}$, calculated. The resulting $\bj$
values are given in column 5 of Table 1. Seven candidate quasars
do not appear in the SSS scans and one object has an $I_{\rm PEAK}$ value
such that saturation effects preclude the calculation of a sensible
magnitude.

\subsection{Notes on individual objects}
\label{section:individual}

\begin{enumerate}
\item{\bf TGN172Z242:} identification based on detection of a broad emission
feature at $4100$\AA\ which may well be spurious.
\item{\bf TGN200Z277:} identification based on identification of $4200$\AA\
feature as \hbox{Mg\,{\sc ii}}
$\lambda$2798. Any H$\beta$ emission would be difficult
to detect because of a fibre absorption band affecting wavelengths
$7150$--$7400$\AA.
\item{\bf TGS306Z029:} redshift, based on \hbox{[O\,{\sc iii}]} 
emission lines at $\sim
6800$\AA\, is secure. Tentative quasar classification relies on
detection of broad \hbox{Mg\,{\sc ii}}
$\lambda$2798 emission at $3800$\AA\ but there
is no evidence for the presence of broad H$\beta$.
\item{\bf TGS522Z221:} unexplained emission feature exists in the
spectrum at 7340\AA.  A higher signal--to--noise ratio 
spectrum is desirable in order to confirm the redshift identification.
\item{\bf TGS187Z041} a spectrum 
showing evidence for a low-redshift galaxy and a quasar at
much higher redshift (see Table.~\ref{table2} and Fig.~\ref{lens}).  
The quasar identification
relies on the presence of a single emission line at $5200$\AA \ and the
redshift of $z=0.87$ is derived assuming the feature is \hbox{Mg\,{\sc ii}}
$\lambda$2798. 
Note that this spectrum was observed using the science-grade CCD1,
whereas it was the engineering-grade CCD2 that produced 
spurious features at $\lambda \simeq 5200$\AA. 
None the less, confirmation of the spectrum is required to establish
TGS187Z041 as a strong lens candidate, particularly since
ESO NTT $V$-band imaging obtained on February 2 2002,
using the SUSI instrument in conditions with $\simeq 1\,$arcsec seeing,
showed no evidence of a morphology indicating the system is a
gravitational lens.
\end{enumerate}

\begin{table*}
 \begin{minipage}{105mm}
 \caption{Candidate quasars found in the 2dFGRS 100k Data Release. Objects
are ordered by their 2dF object name.  RA is given in hours and
 Declination in degrees.  The 2dF and corrected
 magnitudes are described in Section~5.2, missing magnitudes  
 (-) could not be calculated from the current SuperCOSMOS scans.  The
 redshift $z_{\rm qso}$ is a best estimate redshift which has been
 confirmed by eye.}
 \begin{tabular}{@{}ccrcclc}
 \hline 
  2dF Name & RA (J2000) & Dec (J2000) & $b_{\rm J(2dF)}$ &
 $b_{\rm J(corr)}$ & $z_{\rm qso}$ & $ID$ \\
 \hline
TGN120Z045 & 12 07 47.71 & $-$03 51 27.4 & 19.34 & 19.26
& 0.35 & 2 \\
TGN124Z166 & 12 26 47.75 & $-$03 25 22.6 & 19.17 & 19.35
& 0.30 & 1 \\
TGN148Z045 & 14 38 59.94 & $-$04 38 35.0 & 18.66 & 19.01
& 1.82 & 3 \\
TGN160Z226 & 10 39 24.66 & $-$03 48 31.6 & 18.37 & 18.70
& 0.79 & 2 \\
TGN170Z040 & 11 30 45.88 & $-$03 46 30.3 & 19.25 & 19.27
& 0.65 & 2 \\
TGN171Z248 & 11 34 09.16 & $-$04 26 53.4 & 19.27 & 19.08
& 0.93 & 2 \\
TGN172Z242 & 11 35 34.97 & $-$03 44 01.3 & 17.36 &   -
& 1.16 & 1 \\
TGN172Z270 & 11 34 33.45 & $-$03 06 55.5 & 19.15 & 19.18
& 0.45 & 3 \\
TGN176Z251 & 11 53 38.86 & $-$04 19 54.0 & 19.08 & 19.70
& 3.41 & 3 \\
TGN181Z175 & 12 20 18.37 & $-$02 41 01.9 & 18.93 & 18.42
& 0.44 & 3 \\
TGN200Z277 & 13 37 21.90 & $-$02 25 12.1 & 19.02 & 19.16
& 0.51 & 2 \\
TGN207Z146 & 14 16 37.88 & $-$03 01 32.1 & 19.37 & 19.36
& 1.66 & 3 \\
TGN224Z190 & 10 37 02.33 & $-$01 49 53.2 & 18.42 & 17.89
& 0.33 & 3 \\
TGN254Z003 & 12 52 20.45 & $-$00 39 49.9 & 19.44 & 19.31
& 1.84 & 3 \\
TGN280Z141 & 14 36 24.82 & $-$00 29 05.0 & 18.82 & 17.88
& 0.32 & 3 \\
TGN288Z181 & 10 14 14.10 & $-$01 59 18.1 & 18.71 & 18.38
& 2.23 & 3 \\
TGN292Z093 & 10 29 35.17 & $-$01 21 39.8 & 18.52 & 17.84
& 1.06 & 3 \\
TGN305Z146 & 11 25 28.45 & $-$01 51 29.8 & 18.97 & 18.63
& 1.37 & 3 \\
TGN338Z080 & 13 57 00.49 &  00 26 28.5 & 18.69 & 20.10
& 1.54 & 3  \\
TGN346Z197 & 14 34 52.70 & $-$00 28 28.1 & 19.35 & 18.70
& 2.24 & 3 \\
TGN367Z176 & 11 01 05.44 & $-$00 01 55.9 & 19.20 &   -
& 2.47 & 3 \\
TGN423Z040 & 10 19 12.96 &  01 29 36.7 & 19.33 &   -
& 1.28 & 3 \\
TGS059Z150 & 21 59 55.98 & $-$24 16 36.0 & 19.36 & 18.72
& 2.30 & 3 \\
TGS061Z180 & 22 09 19.07 & $-$24 07 13.2 & 19.29 & 18.59
& 0.32 & 3 \\
TGS109Z209 & 03 05 24.31 & $-$24 42 05.5 & 18.77 & 17.71
& 0.51 & 3 \\
TGS114Z079 & 22 00 28.13 & $-$25 19 39.7 & 19.36 & 18.90
& 0.90 & 2 \\
TGS116Z170 & 22 10 44.87 & $-$26 06 25.0 & 19.23 & 18.64
& 0.41 & 3 \\
TGS120Z117 & 22 33 57.46 & $-$25 40 47.0 & 19.39 & 19.58
& 3.03 & 3 \\
TGS132Z079 & 23 53 34.35 & $-$25 52 15.7 & 19.20 & 20.49
& 1.07 & 2 \\
TGS164Z031 & 03 07 43.75 & $-$25 49 03.9 & 19.35 & 19.48
& 1.65 & 3 \\
TGS176Z118 & 22 12 18.36 & $-$27 42 12.2 & 19.19 & 19.27
& 0.53 & 3 \\
TGS185Z206 & 23 00 37.69 & $-$26 44 36.1 & 19.20 & 19.39
& 1.75 & 3 \\
TGS201Z045 & 00 11 28.67 & $-$28 11 23.2 & 19.42 &   -
& 2.27 & 3 \\
TGS215Z047 & 01 20 17.37 & $-$27 45 10.8 & 19.20 & 18.69
& 2.57 & 3 \\
TGS216Z071 & 01 18 56.85 & $-$27 26 33.4 & 19.05 & 18.44
& 1.80 & 3 \\
TGS296Z199 & 01 18 09.10 & $-$29 01 54.1 & 19.30 & 20.33
& 1.73 & 3 \\
TGS299Z124 & 01 31 32.12 & $-$28 33 09.9 & 19.32 &   -
& 1.12 & 3 \\
TGS306Z029 & 02 08 14.01 & $-$28 28 10.6 & 19.12 & 20.22
& 0.36 & 1 \\
TGS322Z050 & 03 31 51.85 & $-$29 00 17.3 & 19.16 &   -
& 0.31 & 1 \\
TGS328Z215 & 21 46 45.76 & $-$30 43 39.6 & 19.33 & 18.78
& 2.63 & 3 \\
TGS340Z134 & 22 38 41.00 & $-$30 51 29.5 & 19.34 & 19.09
& 1.65 & 3 \\
TGS391Z057 & 02 28 12.58 & $-$30 47 49.9 & 19.28 & 18.20
& 0.91 & 2 \\
TGS393Z082 & 02 45 00.78 & $-$30 07 23.0 & 19.34 & 18.28
& 0.34 & 3 \\
TGS467Z204 & 02 48 24.66 & $-$31 33 48.9 & 18.50 & 17.19
& 0.32 & 3 \\
TGS468Z111 & 02 59 10.91 & $-$31 40 08.4 & 19.25 & 18.79
& 0.43 & 3 \\
TGS522Z221 & 02 55 22.78 & $-$32 45 55.4 & 18.92 & 19.58
& 2.73 & 3 \\
TGS817Z145 & 03 57 03.57 & $-$49 25 40.9 & 19.44 & 19.25
& 0.78 & 2 \\
TGS829Z039 & 01 30 11.32 & $-$47 59 55.9 & 19.43 & 20.03
& 2.45 & 3 \\
TGS829Z217 & 01 24 27.01 & $-$47 31 30.5 & 19.25 & 18.61
& 1.56 & 3 \\
TGS836Z293 & 01 17 29.50 & $-$39 14 19.3 & 19.09 & 19.64
& 2.11 & 3 \\
TGS836Z539 & 01 20 00.31 & $-$40 56 59.3 & 19.11 & 18.43
& 1.18 & 3 \\
TGS858Z005 & 00 24 35.98 & $-$12 32 03.4 & 19.41 & 19.80
& 2.47 & 3 \\
TGS875Z416 & 02 38 58.98 & $-$46 06 01.6 & 19.27 & 18.67
& 1.14 & 3 \\
TGS893Z369 & 00 23 20.07 & $-$40 34 18.5 & 19.38 & 18.71
& 2.63 & 3 \\
TGS901Z004 & 01 22 29.47 & $-$08 30 11.8 & 19.03 &   -
& 1.93 & 1 \\
  \hline
 \end{tabular}
 \label{table1}
\end{minipage}
\end{table*}

\begin{table*}
 \begin{minipage}{105mm}
 \caption{Candidate composite galaxy plus quasar spectrum.  Listed are
 the 2dF and corrected $\bj$ magnitudes as well as the estimated
 redshifts of the galaxy and quasar components.}
 \begin{tabular}{@{}cccccccc}
 \hline
  2dF Name & RA (J2000) & Dec (J2000) & $b_{\rm J(2dF)}$ &
 $b_{\rm J(corr)}$ & $z_{\rm gal}$ & $z_{\rm qso}$\\
 \hline
TGS187Z041 & 23 11 40.62 & $-$27 47 44.6 & 17.94 & 16.83 &
 0.1643 & 0.87\\
  \hline
 \end{tabular}
 \label{table2}
\end{minipage}
\end{table*}

\section{Discussion}
\label{section:discussion}

The sample of high-redshift quasars described in 
Section~\ref{section:catalogue} is unique due to the parent
survey's explicit morphological selection of non-stellar objects.
Indeed, most of the images shown in Fig.~\ref{sample} 
are either noticeably extended or have distinct companions. 
From a more physical point of view 
the quasars fall into three distinct categories:
those with visible host galaxies (Section~\ref{section:host});
possible gravitational lenses (Section~\ref{section:lens});
and 
those with low-redshift galaxies along the line--of--sight 
(Section~\ref{section:companion}).

\subsection{Quasar host galaxies}
\label{section:host}

A significant proportion of the quasars in the sample (13 out of
55) have redshifts $z \la 0.5$ and sufficiently non-stellar appearances
to have entered the APM survey without any nearby companions 
(\cf\ the compound objects described in
Sections~\ref{section:lens} and \ref{section:companion}).
In several cases there is 
spectroscopic evidence for the presence of host galaxies
in the form of common photospheric absorption lines,
implying that it is the quasars' hosts that are also responsible 
for the extended morphology. 
Although the discovery of this kind of low-redshift quasar was not a strong
motivation for this search, imaging and spectroscopy of these 
objects, which would be difficult to find in any other way, 
could prove valuable in the understanding of the relationship between
quasars and their host galaxies.

\subsection{Gravitational lenses}
\label{section:lens}

\begin{figure}
 \psfig{figure=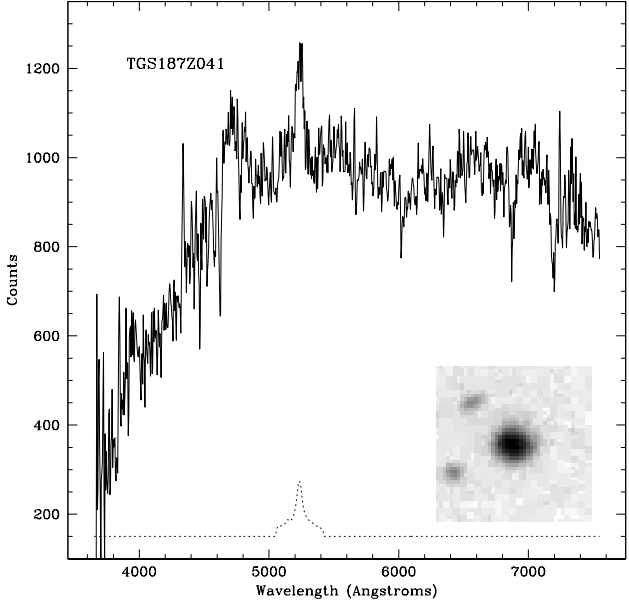,width=3.5in}
 \caption{Image and spectrum of TGS187Z041.  This object displays
  both a galaxy ($z\simeq 0.16$) and a quasar ($z\simeq 0.87$)
 component in its spectrum.  The image is a $30\times 30$ arcsec
 postage stamp taken from the SuperCOSMOS Sky Survey.  Also shown
 (dotted line) is the corresponding quasar template \hbox{Mg\,{\sc ii}}
 emission feature.}
 \label{lens}
\end{figure}

One of the objects discovered in this survey,
TGS187Z041 (Section~\ref{section:individual}; Fig.~\ref{lens}),
is a possible gravitational lens, having the morphology of a low-redshift
galaxy, but a spectrum that appears to be a combination of galaxy
and quasar. Although higher resolution imaging is required to 
confirm the lensing hypothesis, such a discovery would not be 
entirely unexpected, and indeed the initial motivation for this project
was to search for spectroscopic lenses (Mortlock, Madgwick \& Lahav 2001). 

To date Q~2237+0305 (Huchra \etal\ 1985) is the only quasar 
lens identified in this fashion\footnote{Related 
surveys for lensed emission line galaxies 
(Warren \etal\ 1996; Hewett \etal\ 2000; Hall \etal\ 2000) have 
resulted in a number of candidates, but only the first of these 
is confirmed at present.},
but theoretical calculations 
by Kochanek (1992) and Mortlock \& Webster (2001, 2002)
implied that $\sim 10$ lenses could be discovered in the 
(full) 2dFGRS. 
Thus $\sim 4$ would be expected in the 100k sample, which, given
that the discovery of these systems is a
Poisson process, is marginally consistent with the single
(putative) lens identified. However the nature of the other 
quasars discovered during the search allows the models used in the above 
calculations to be reassessed. 

In particular, the expected number of lenses increases rapidly with 
$\Delta m_{\rm qg}$, the parameter which specifies how much fainter 
than the galaxy the quasar images can be whilst still being detectable.
Mortlock \& Webster (2001) estimated $\Delta m_{\rm qg} \simeq 2$ for
the 2dFGRS, but the absence of galaxy-dominated spectra 
from the objects described in Section~\ref{section:host}, 
together with Monte Carlo simulations, implies that a 
value of $\Delta m_{\rm qg} \simeq 1$ is more appropriate.
It is also clear in some cases
that the identification of the quasar emission lines
is limited by the photon noise as well as the presence of 
galactic emission, a factor explicitly accounted for by neither
Kochanek (1992) nor Mortlock \& Webster (2000, 2001).
The estimated SSS magnitudes of the high-redshift objects in Table~\ref{table1}
imply an effective limiting magnitude of $\bj \simeq 20$ for
the quasars themselves, which would certainly remove some lenses
from the sample if $\Delta m_{\rm qg} \simeq 2$. However the adjustment
to $\Delta m_{\rm qg} \simeq 1$ also has the effect of excluding 
noise-limited detections, implying that the above models are 
adequate provided a value of $\Delta m_{\rm qg}$ close to unity is adopted.
Overall it seems that the results are consistent with this more 
conservative model of the 2dFGRS, which then implies a total yield of 
$\sim 3$ lenses from the full survey.

\subsection{Absorption systems}
\label{section:companion}

Most of the redshift $z \ga 0.5$ quasars in the sample entered
the 2dFGRS due to the presence of faint companion images --
some classes of `star--galaxy' mergers were selected into the parent
APM survey (Maddox \etal\ 1990). In most cases the companions 
appear to be faint ($\bj \simeq 19$--$21$) low-redshift galaxies,
although this is in part due to the strong selection effect that the 
quasars be detectable in the 2dF spectra.
(If there is significant angular
separation between the quasar its companion this requires that the
2dF fibre was centred close to the quasar;
in situations where the two components are
coincident the quasar must be approximately as bright as its companion
for its emission lines to be detectable; \cf\ Section~\ref{section:lens}.)

Confirmation imaging and spectroscopy of the quasar candidates with
companion images should then provide a well characterised sample
of redshift $\la 0.3$ galaxies that possess background probes (the quasars)
with impact parameters ranging from $\sim 2$ to $\sim 30\,$kpc.
Such systems will be of value in studies of the interstellar medium of
the foreground galaxies and for establishing the properties of
inter-galactic absorption at low redshift (\eg\ Bowen, Blades \& Tytler
1997; Bowen, Blades \& Pettini 1995; Chen \etal\ 1998, 2001).

\section{Conclusions}
\label{section:conclusion}

We have conducted a spectroscopic search for quasars in the 
2dFGRS 100k catalogue and generated a sample of 55 candidate
objects. Two analysis techniques were used -- a matched filter-based
search and wavelet transforms -- and both were successful in finding
$\sim 90$ per cent of the candidates. 
The low-redshift quasars in the sample are typically extended 
(probably due to their host galaxies) and many of the high-redshift 
quasars have fainter companions (likely to be low redshift galaxies
along the line--of--sight), including one which is a possible gravitational
lens. We are currently in the process of obtaining 
confirmation spectroscopy of these objects (and imaging of the
candidate lens). 
The next stage of this project will be to obtain high quality 
spectra of the galaxies with good background probes 
(\ie\ the quasars with low impact parameters) with a view 
to performing absorption studies of their interstellar medium.

\begin{figure*}
 \psfig{figure=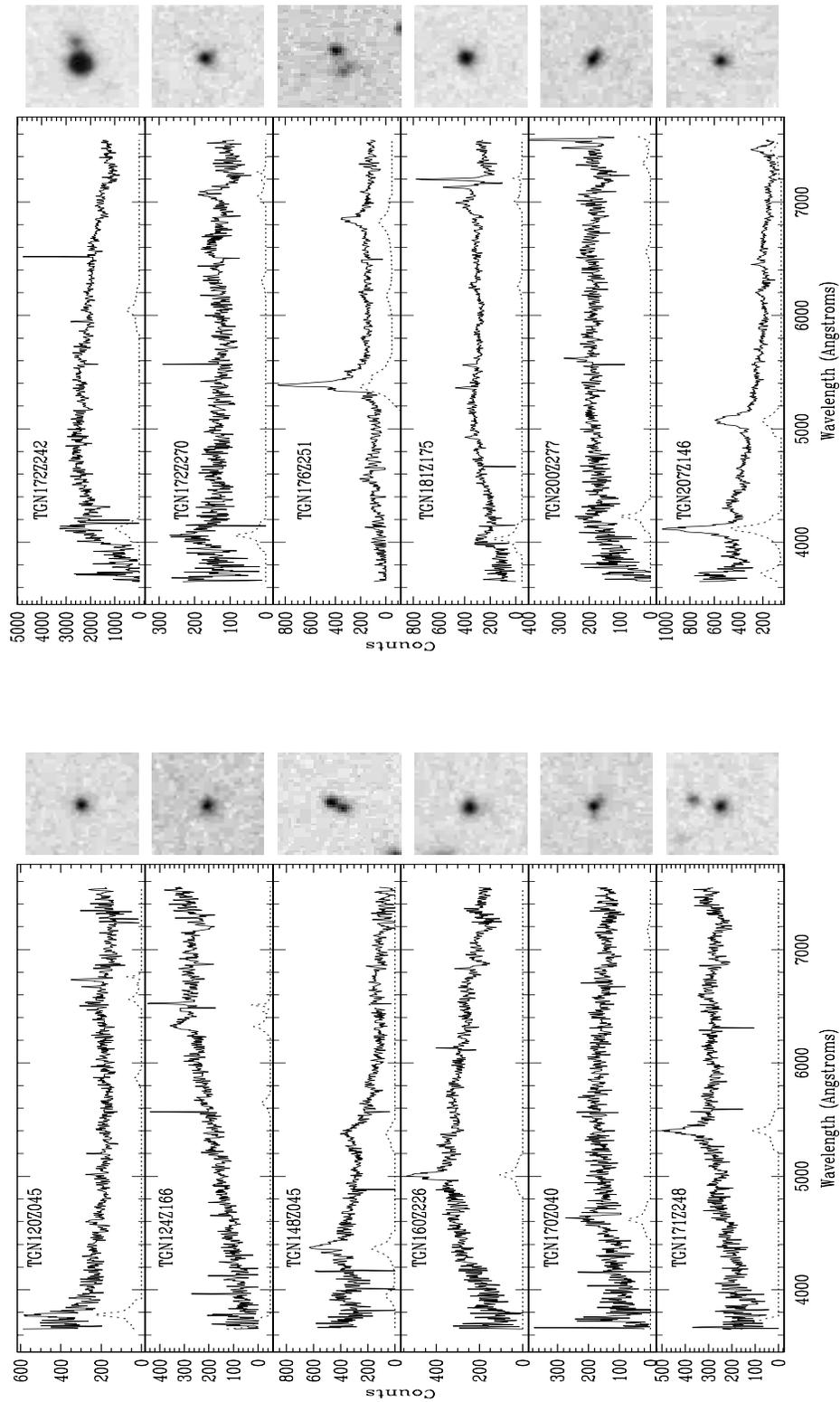,width=7in,height=9in}
 \caption{Images and spectra of identified high-redshift quasars in
 the 2dFGRS 100k Data Release.  The displayed images are $30\times 30$ arcsec
 postage stamps taken from the SSS.  Also shown
 (dotted line) are the redshifted quasar emission line templates used
 to identify each object.}
 \label{sample}
\end{figure*}

\begin{figure*}
 \psfig{figure=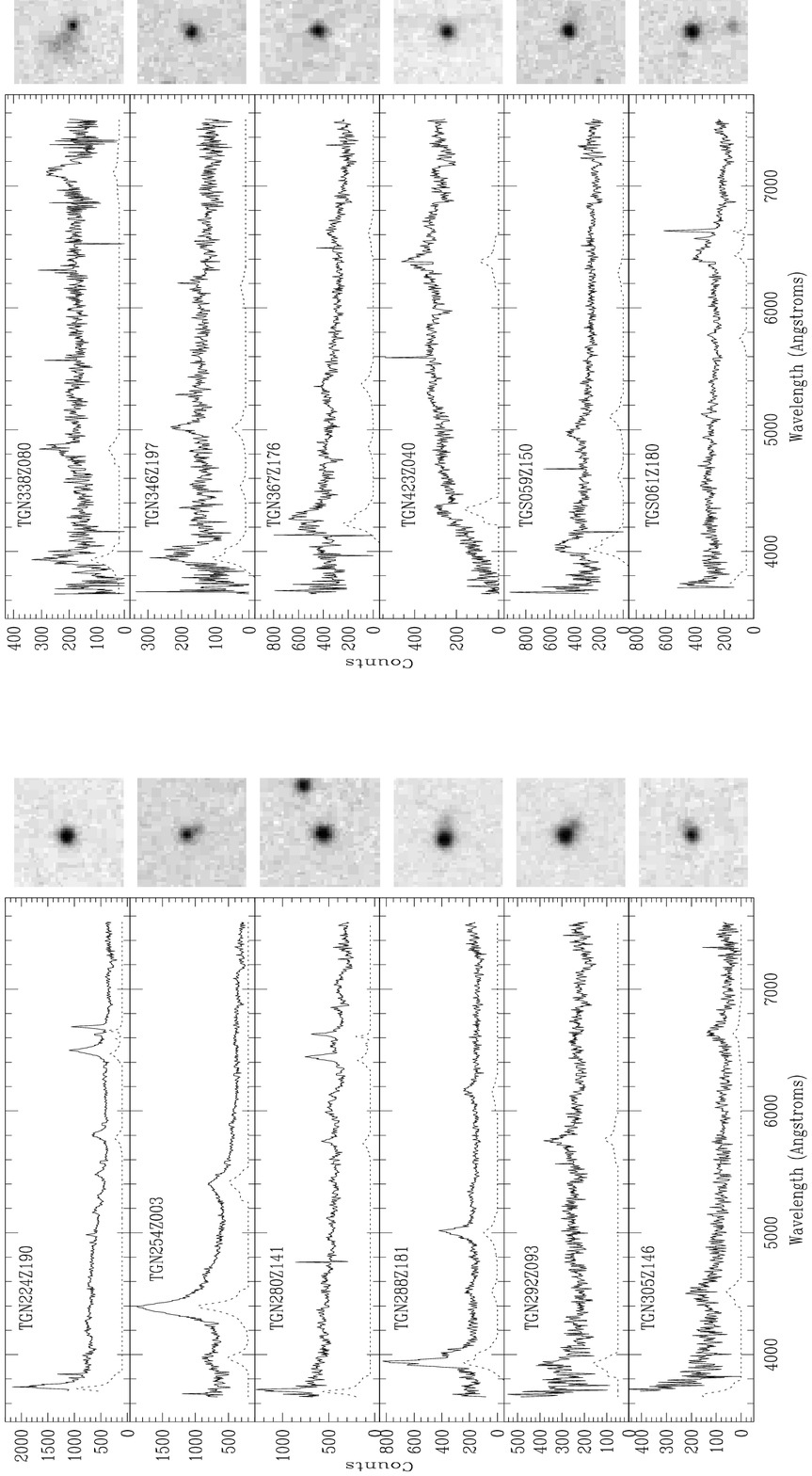,width=7in,height=9in}
 \contcaption{}
\end{figure*}

\begin{figure*}
 \psfig{figure=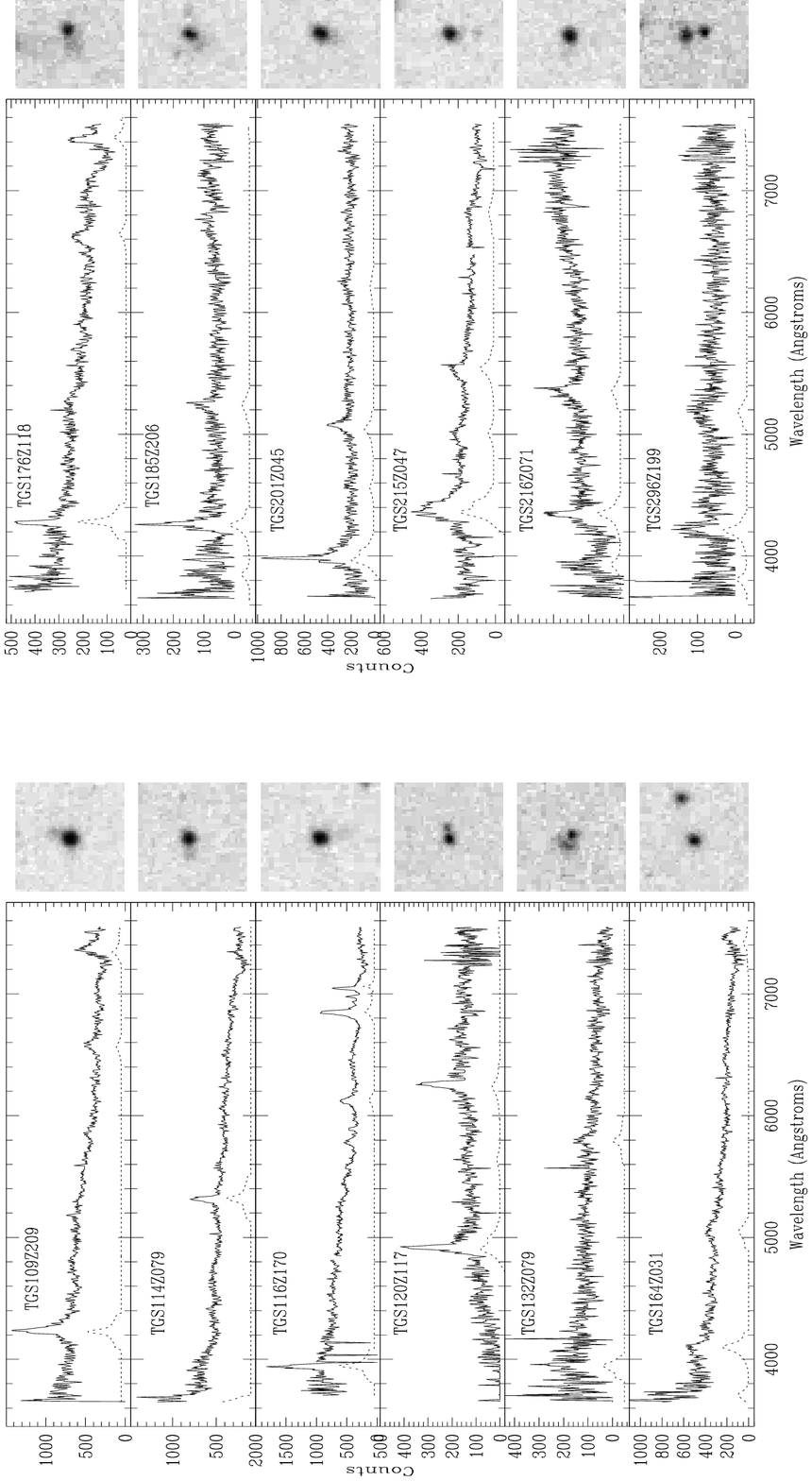,width=7in,height=9in}
 \contcaption{}
\end{figure*}

\begin{figure*}
 \psfig{figure=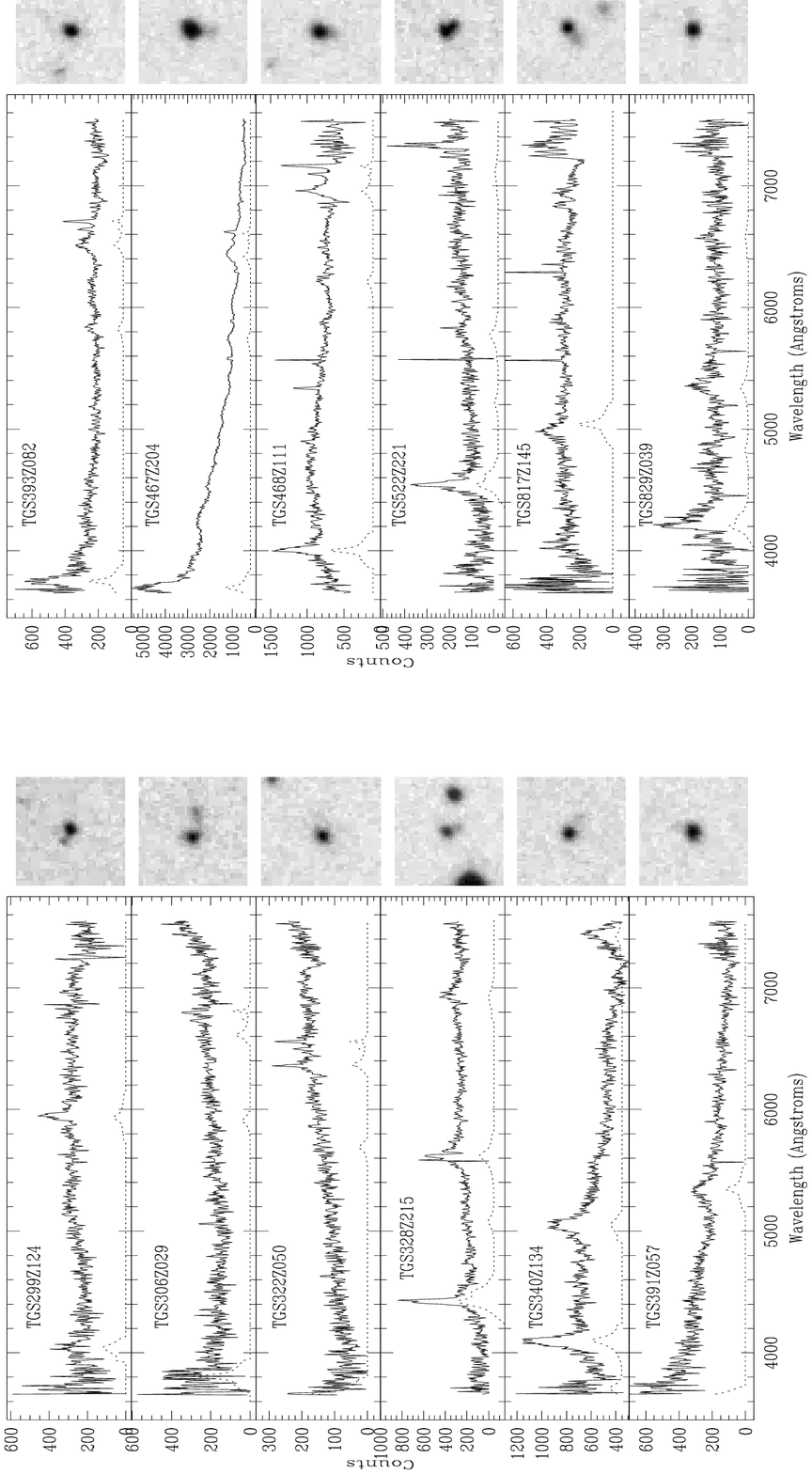,width=7in,height=9in}
 \contcaption{}
\end{figure*}

\begin{figure*}
 \psfig{figure=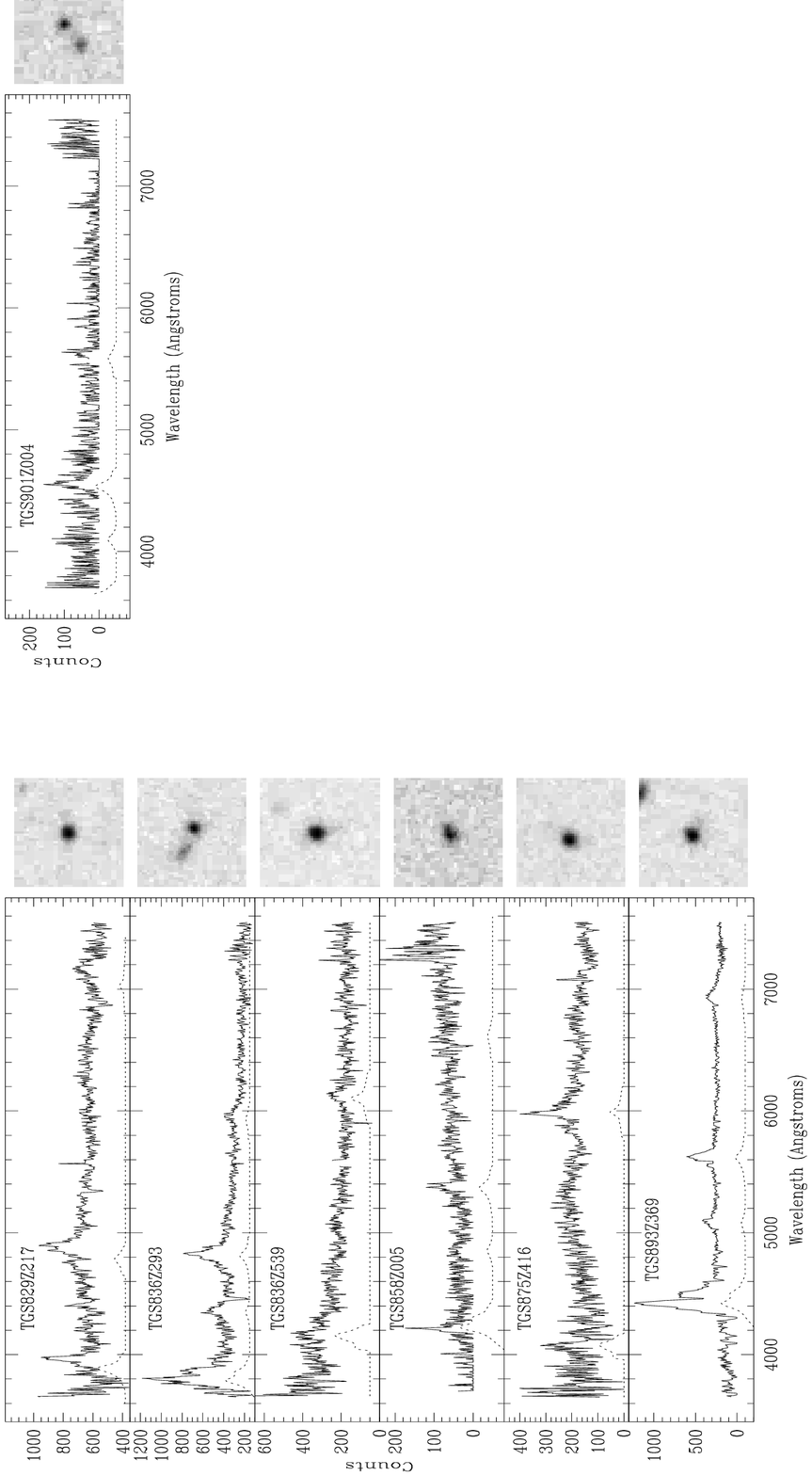,width=7in,height=9in}
 \contcaption{}
\end{figure*}

\section*{Acknowledgments}

The efforts of the 2dFGRS team in compiling the 100k Data Release are
greatly appreciated. Terry Bridges, Russell Cannon and Karl Glazebrook
provided valuable
advice concerning features of the 2dFGRS spectra.  In addition Russell
Cannon also provided us with many valuable comments on improving the
original draft of this paper.
Tom Oosterloo and
Lister Staveley--Smith contributed significantly to our understanding of
wavelets and Mike Read provided invaluable assistance with the use of the
SuperCOSMOS Sky Survey.  DSM was supported by
an Isaac Newton Studentship from Trinity College and the University of
Cambridge. DJM was supported by PPARC.

\bsp
\label{lastpage}
\end{document}